\documentstyle[preprint,aps]{revtex}
\newcommand{\referencestyle}{
\small
\abovedisplayskip=6pt
\belowdisplayskip=6pt
\vspace{12pt}}

\def\De{\Delta\eta}
\def\Det{\Delta\eta_T}

\def\Dt{\Delta\tau}
\def\Dy{\Delta y}
\def\tl{\tau}
\def\t{{\tau}}
\def\ch{\cosh}
\def\sh{\sinh}
\def\ben{\begin{eqnarray}}
\def\enn{\end{eqnarray}}
\def\ov{\over\displaystyle\strut}
\def\dst{\displaystyle\strut}
\def\l({\left(}
\def\r){\right)}
\def\o{{out}}
\def\s{{side}}
\def\e{{\eta}}

\begin{document}
\include{epsf}
\rightline{CU-TP-717, hep-ph/9509213}
\rightline{LUNFD6/(NFFL-7082)-Rev. 1994}
\vfill

\begin{center}
{ \large\bf
Bose-Einstein Correlations for Three-Dimensionally Expanding\\
Cylindrically Symmetric Finite Systems
}
\end{center}
\medskip
\medskip
\begin{center}
	T. Cs\"org\H o$^{1,2}$
	 and    B. L\"orstad$^{3}$
	\footnote{
	csorgo@sunserv.kfki.hu,~
	bengt@quark.lu.se}
\end{center}


\begin{center}
{\it
$^1$Department of Physics, Columbia University,
 New York, NY  10027, USA, \\
$^2$MTA KFKI RMKI,
 H--1525 Budapest 114, P.O. Box 49. Hungary, \\
$^3$Physics Institute,
 University of Lund, Professorsgatan 1, S - 223 63 Lund, Sweden \\
}
\end{center}
\vfill


\date{September 26, 1995}


\begin{abstract}
There are {\it two type of scales} present simultaneously
in the space-like as well as in the time-like directions in
a model-class describing a cylindrically symmetric, finite,
three-dimensionally expanding boson source.
One type of the scales is related to the finite lifetime or geometrical size
of the system, the other type is governed by the rate of
change of the local momentum distribution
in the considered temporal or spatial direction.
The parameters of the Bose-Einstein correlation function may
obey  an {\it $M_t$-scaling}, as observed in $S + Pb$ and $Pb + Pb$
reactions at CERN SPS.
  This $M_t$-scaling may imply that the
 Bose-Einstein correlation functions view only a small part of a
 big and expanding system. The full sizes of the
expanding system at the last interaction are shown to be measurable
with the help the invariant momentum distribution of the emitted particles.
A vanishing duration parameter
can also be generated, with a specific $M_t$ dependence,
in the considered model-class.
\end{abstract}

 \vfill\eject
{\section{Introduction}}
The method of intensity interferometry has recently become a
widely used tool for determining the space-time picture of
high energy heavy ion collisions. Originally, the method was
invented~\cite{HBT} to measure angular diameters of distant stars.
The objects under study were approximately static and the length-scales
astronomical. In principle the same method is applied
to measure space-time characteristics of
high energy heavy ion collisions, where the objects are expanding
systems, with life-times of a few fm/c ($10^{-23}$ sec)
and length-scales of a few fm ($10^{-15}$ m).

In the case of high energy heavy-ion collisions intensity interferometry
is pursued  to infer the equation of state
and identify the possible formation of a
transient Quark-Gluon Plasma state from a determination
of the freeze-out hyper-surface, as scanned by the Bose-Einstein
correlation function (BECF), see e. g. the contributions of
the NA35, NA44 and WA80 collaborations in ref.~\cite{QM,na44}.
For  introduction and review on Bose-Einstein
 correlations  see refs.~\cite{bengt,zajc}.
Non-trivial effects arising from correlations among space-time
and momentum-space variables were studied in refs.~\cite{padu}.

The recent $^{32}S + ~ ^{197}Pb$ reactions at 200 AGeV laboratory
bombarding energy resulted in a non-expected, symmetrical BECF-s~\cite{na44}
if measured in the LCMS, the longitudinally comoving
system of the boson pairs~\cite{lutp}.
The longitudinal radius parameter was shown to measure a
length-scale,
$R_L \propto 1/\sqrt{m_t}$, introduced in ref.~\cite{sinyukov}
for an infinite, longitudinally expanding Bjorken tube.
The {\it side} radius parameter was thought to measure the geometrical
radius and the {\it out} component to be sensitive to the
duration of the particle freeze-out times~\cite{lutp,1d}.
All radius component parameters turned out to be equal within the experimental
errors.
Although {\it this might be just a coincidence},
 in this work we show that
such a behavior, valid in a certain $m_t$ interval, {\it may also be a
natural consequence of a cylindrically symmetric three-dimensional
hydrodynamic expansion}. In this case the local temperature,
the gradients of the temperature distribution
and the flow-gradients
generate `thermal' length-scales in all these space-like directions.
Changes in the local temperature during the particle emission
induce a temporal scale, the thermal duration.
Recently it became clear that the parameters of the
BECF-s measure the {\it lengths of
 homogeneity}~\cite{sinyukov,sinyu93,hhm:te,uli} which in turn were shown
to be expressible in terms of the geometrical and the thermal lengths
{}~\cite{nr,1d,uli}.

We shall derive here general relationships
among the functional forms of the BECF-s as given in the
laboratory (LAB) frame, the LCMS frame and the
longitudinal saddle-point system (LSPS) in which
the functional form of the BECF-s turns out to be the simplest one.

A new class of analytically solvable models is introduced thereafter,
describing a three-dimensionally expanding, cylindrically
symmetric system
for which the geometrical sizes and the duration
of the particle emission are finite.
In this class of the models there are two length-scales
present in all directions, including the temporal one.
The BECF is found to be dominated by the shorter,
while the momentum distribution by the
longer of these scales.
The interplay between the finite "geometrical scales" of the
boson-emitting source and the finite "thermal scales"
shall be considered in detail.

\medskip\medskip
\section{Formalism}
Both the momentum spectra and
the BECF-s are prescribed
in the applied Wigner-function formalism~\cite{pratt_csorgo,bengt}.
In this formalism the BECF is calculated from the two-body Wigner-function
assuming chaotic particle emission. In the final expression the
time-derivative of the (non-relativistic) Wigner function
is approximated~\cite{lutp,pratt_csorgo}
 by a classical emission function $S(x;p)$,
which is the probability that a boson is produced
at a given $ x = (t, {\bf r} \,) = (t,r_x,r_y,r_z)$ point in space-time
with the four-momentum $p = (E, {\bf p}\,) = (E, p_x, p_y, p_z)$.
The emission function has been related to the covariant Wigner-transform
of the density matrix of pion sources in refs.~\cite{pratt_csorgo,zajc}
and  most recently in ref.~\cite{swig}, where  the relation of Wigner-function
formalism to the
covariant current formalism ~\cite{gyulassy} has also been clarified.
The (off-shell) two-particle Wigner functions shall be approximated by the
off-shell
continuation of the on-shell Wigner-functions~\cite{pratt_csorgo,nr,1d}.
The particle is on the mass shell, $ m^2 = E^2 - {\bf p}^{ 2} $.
Please note the difference between $x$ indicating a four-vector
in space-time
and the script-size $_x$ which indexes a direction in coordinate space.

A useful auxiliary function is the Fourier-transformed
emission function
\ben
\tilde S(\Delta k ; K ) & = & \int d^4 x \, S(x; K) \,
	\exp(i \Delta k \cdot x ),
\enn
where
\begin{eqnarray}
\Delta k  = p_1 - p_2, & \qquad\qquad\qquad & K  = {\displaystyle\strut p_1 +
p_2 \ov 2}
\end{eqnarray}
and $\Delta k \cdot x $ stands for the inner-product
of the four-vectors.
Then the one-particle inclusive invariant momentum distribution
(IMD) of the emitted particles,
$N_1({\bf p})$ is given by
\ben
N_1({\bf p}) & = & \tilde S(\Delta k = 0; K = p) \,
	 = {\dst E \ov \sigma_{tot}}  {\dst d\sigma \ov d{\bf p}},
	\label{e:imdd}
\enn
where $\sigma_{tot}$ is the total inelastic cross-section.
This IMD is normalized to the mean multiplicity $\langle n \rangle $ as
\ben
\int {d{\bf p} \ov \dst E} N_1({\bf p}) & = & \langle n \rangle.
 \label{e:nor}
\enn
In the present paper effects arising from the final state  Coulomb and Yukawa
interactions
shall be neglected.
The two-particle BECF can be calculated
from the emission function with the help of the
well-established  approximation
\ben
C(\Delta k;K) & = & {\dst \langle n\rangle^2 \ov \langle n(n- 1)  \rangle}
		{\dst N_2 ({\bf p}_1,{\bf p}_2)
		\ov N_1({\bf p}_1) \, N_1({\bf p}_2) }
		 \simeq
		1 + {\displaystyle\strut \mid \tilde S(\Delta k ; K) \mid^2
				\ov
			\mid \tilde S(0;K)\mid^2 },
\label{e:c2gen}
\enn
utilized also in ref.~\cite{nr},
see ref. \cite{pratt_csorgo} for further details.
The corrections to this expression are known to be small~\cite{uli}.
Note that among the eight components of $\Delta k $ and $K$ only six
are independent due to the two constraints $ p_1^2 = p_2^2 = m^2$.
These constraints can be formulated alternatively as
$ \Delta k \cdot K = 0$ and  $ K^2 = m^2 - \Delta k^2 /4$. Thus the
two-particle BECF depends on the off-shell emission function, which we
approximate by the off-shell continuation of the on-shell emission
functions.

 A similar but not identical approximation used by several authors
 is to replace
$\tilde S(\Delta k; K)$ by $\tilde S(\Delta k; K')$ where
off-shell $K$ is changed to   on-shell $K'$. The latter mean momentum is
defined to be on-shell as
$K'^{0} = m^2 - {\bf K'}^2 $ where ${\bf K'} = {\bf K}
 = ({\bf p_1} + {\bf p_2}) /2$. The differences
between these two approximation schemes are of ${\cal O} (\Delta k^2 /m^2)$.
The above two approximation schemes coincide in the $\Delta k^2  \rightarrow 0$
limit
where the Bose-Einstein correlations are maximal.
Since we shall make use of the $ \Delta k \cdot K = 0$ constraint
which is exact only if the $K$ four-vector is off-shell, we shall
approximate the off-shell emission function in  eq. (~\ref{e:c2gen})
 with the off-shell continuation
of the on-shell emission function.

\section{General considerations}
We model the emission function in terms of the longitudinally boost-invariant
variables:  the (longitudinal) proper-time,
$\tau= \sqrt{t^2 - r_z^2}$, the
space-time rapidity $\eta = 0.5 \ln\l(\,(t+r_z)\,/\,( t-r_z)\, \r)$,
the transverse mass $m_t = \sqrt{E^2 - p_z^2}$
 and the momentum-space rapidity
$\mbox{\rm y} = 0.5 \ln\l(\,(E + p_z )\,/\,(E - p_z)\, \r)$.
  In the transverse direction, the transverse radius,
 $r_t = \sqrt{r_x^2 + r_y^2}$ is introduced.
We have
\ben
	t = \tl \ch(\eta), & \qquad\qquad\qquad & r_z = \tl \sh(\eta).
\enn
For systems undergoing a boost-invariant longitudinal expansion,
the emission function may be a function of
boost-invariant variables only. These are
$\tau$, $r_x,r_y$, $p_x,p_y$ and
$\eta - y$.
However, for finite systems the exact longitudinal boost-invariance cannot
be achieved and the emission function becomes a function of $\eta - y_0$
too, where $y_0$ stands for the mid-rapidity.
Approximate boost-invariance is recovered  in the mid-rapidity
region only, where terms proportional to $\eta - y_0$ can be neglected.
Thus for finite systems undergoing a boost-invariant longitudinal expansion
the emission function can be given in terms of
these variables as
\ben
	S(x;K) \, d^4x & = & S_*(\tau,\eta,r_x,r_y) \,
				d\tau \, \tau_0 d\eta \, dr_x \, dr_y.
\enn
Here we introduced the constant $\tau_0$ in front of $d\eta $ due to
dimensional
reasons and included the Jacobian from the $d^4 x $ to the
$d\tau \,  d\eta \, dr_x \, dr_y$ variables
into the emission function $S_*(\tau,\eta,r_x,r_y)$.
The subscript $_*$ indicates that the functional form of the emission function
is changed with the change of the variables.
Further,  dependences on the mean momentum $K$ as well as on the
mid-rapidity $y_0$ are also indicated with
the subscript $_*$.  The effective, momentum-dependent parameters of the
emission function $S_*(\tau,\eta, r_x, r_y)$ shall also be indexed with
$_*$ in the forthcoming.
The subscript $_s$ stands for the
point where the emission function is maximal (we assume that $S(x;K)$
has only one maximum for any values of $K$). We do not assume at this point
 whether the function $- \ln S(x;K)$
is expandable into a (multi-variate) Taylor series~\cite{uli}
 around its unique minimum at
the saddle point $x_s$ or not,
merely we assume that the Fourier-transformed $\tilde S(\Delta k;K)$ exists.
See the Appendix for a clarifying example.
We suppose, however,
 that the Fourier-transformed $\tilde S(\Delta k;K)$
can be evaluated in terms of the $\tau $ and $\eta$ variables
in the small $\Delta k$ region relevant for the analysis of the BECF-s.
This is possible if
 the region around $x_s(K)$, where the Fourier integrals pick up
the dominant contribution from,
is sufficiently small so that within this region
the $\tau$ and $\eta$ dependence of $t$ and $r_z$ can be linearized as
\ben
	t & \simeq & \tau \cosh[\eta_s] +   (\eta - \eta_s) \tau_s \sinh[\eta_s]
		 \label{e:t}, \\
	r_z & \simeq & \tau \sinh[\eta_s] + (\eta - \eta_s) \tau_s \cosh[\eta_s]
		\label{e:z},
\enn
with negligible second-order corrections. This condition is fulfilled
if the characteristic sizes $\Delta\tau_*$ and $ \Delta\eta $
of the considered region around
$x_s(K)$ satisfied  $\Delta\tau_*^2 << \tau_s^2 $ and $\Delta\eta_*^2 << 1$.

The principal directions for the decomposition
of the relative momentum at a given value of
the mean four-momentum $K$ are given as~\cite{bertsch,lutp}:
the {\it \o } direction is parallel to the component of ${\bf K}$,
which is perpendicular to the beam, indexed with $_{\o}$,
the {\it longitudinal} or {\it long}  direction is parallel to the beam-axis
$r_z$,
this component of the relative momentum is indexed with $_L$,
and the remaining direction orthogonal to both {\it longitudinal }
and {\it  \o} is called the {\it \s} direction, indexed with $_{\s}$.
Thus the mean and the relative momenta are decomposed as
$K = (K_0,K_{\o},0,K_L)$ and $ \Delta k = (Q_0,Q_{\o},Q_{side},Q_L)$.

Since the particles are on mass-shell, we have
\ben
0 & = & K \cdot \Delta k = K_0 Q_0 - K_{\o} Q_{\o} - K_L Q_L .
\enn
Thus the energy difference $Q_0$ can be expressed as
\ben
	Q_0 & = &  \beta_{\o} Q_\o + \beta_L Q_L,   \label{e:11}
\enn
where we have introduced the longitudinal and the outward
component of the velocity of the pair, $\beta_L  = K_L / K_0$
and $\beta_\o = K_\o / K_0$, respectively.
These relations become further simplified in the LCMS,
the longitudinally co-moving system, introduced first in ref.~\cite{lutp}.
The LCMS is the frame where $K_L = 0$
thus $\beta_L = 0$. We also have
$\beta_{\o} = \beta_t$ where $_t$ stands for transverse, e.g.
$r_t = \sqrt{r_x^2 + r_y^2}$ and $m_t = \sqrt{m^2 + p_x^2 + p_y^2}$.
Note that the relation $\beta_{\o} = \beta_t $ is independent
of the longitudinal boosts, but both sides of this equation
transform like $1/K_0$.

Let us express the Fourier integrals in terms of the $\tau $ and
$\eta$ variables in the laboratory reference frame (LAB),
utilizing eqs.~(\ref{e:t},\ref{e:z}).
The results in LCMS can be obtained
from the more complicated results in the LAB frame
by the substitution $\beta_L = 0$ and $\beta_{\o} = \beta_t$.
To simplify the notation, let us rewrite
\ben
	\Delta k \cdot x & = & Q_0 t - Q_\o r_x - Q_\s r_y - Q_L r_z \simeq
		Q_\t \t - Q_\o r_x - Q_\s r_y - Q_\e \t_s (\eta - \eta_s),
\enn
utilizing the linearized eqs.~(\ref{e:t},\ref{e:z}). We have introduced the
coefficients of the $\t$ and the $\t_s (\e - \e_s)$ as new variables
given by
\ben
	Q_\t & = & Q_0 \cosh[\eta_s] - Q_L \sinh[\eta_s] =
		(\beta_t Q_\o + \beta_L Q_L)  \cosh[\eta_s] - Q_L \sinh[\eta_s], \\
	Q_\e & = &  Q_L \cosh[\eta_s] - Q_0 \sinh[\eta_s] =
		Q_L \cosh[\eta_s] -
		(\beta_t Q_\o + \beta_L Q_L)  \sinh[\eta_s].
\enn
{}From these relations it follows that
\ben
	C(\Delta k;K) \simeq
		1 + {\displaystyle\strut \mid \tilde S(\Delta k ; K) \mid^2
				\ov
			\mid \tilde S(0;K)\mid^2 }
		\simeq
			1 +
		{\dst \mid \tilde S_*(Q_\t, Q_\e, Q_\o, Q_\s) \mid^2 \ov
			\mid \tilde S_*(0,0,0,0) \mid^2 }.
	\label{e:best}
\enn
Note that this expression
contains a four-dimensional Fourier-transformed function,
and among the four variables $Q_\t,Q_\e, Q_\o$ and $Q_\s$ only three
are independent due to eq.~(\ref{e:11}). Note also that at this point
the BECF may have a non-Gaussian structure, and its dependence on
its variables does not factorize.
The main limitation of the last approximation in eq.~(\ref{e:best})
 is that it is valid only for systems with small lengths of homogeneity,
$\Delta\tau_*^2 << \tau_s^2 $ and $\Delta \eta_*^2 <<1 $.
	As we shall see in the forthcoming, this gives a
	lower limit in $m_t$ for the applicability
	of the simple analytic results for a certain
	class of emission functions.

\section{The core/halo model}
If the system under consideration consists of a {\it core} characterized
by a hydrodynamic expansion and small regions of homogeneity,
and a surrounding {\it halo} of long-lived resonances,
then the above general expression can be further evaluated if
the halo is characterized by sufficiently large regions of homogeneity.
Indeed, the long lived resonances may decay in a large volume proportional
to their lifetime, and the decay products are emitted with a given momentum
distribution from the whole volume of the decay.

The key point is the following:
Let us consider an ensemble of long-lived resonances
with similar momentum, emitted from a given small volume of the core.
The momentum distribution of the {\it decay products} of these resonances
will be similar to each other, independently of the approximate position
of the decay. Now the approximate position of the decay is randomly
distributed along the line of the resonance propagation with the
weight $P(t) \propto \exp(- m_{res} \Gamma_{res} t /E_{res} )$.
Thus the decay products will be emitted with the same momentum
distribution from a volume which is
elongated along the line of resonance propagation, given by
$V_{decay} \simeq A_0 \mid p_{res} \mid /(m_{res} \Gamma_{res})$,
where $A_0$ is the initial transverse size of the surface
through which the resonances are emitted with a momentum $p_{res}$
approximately at the time of the decay of the core, $\tau_s$.

 Thus the halo of long-lived
resonances is characterized by large regions of homogeneity.
(In case of the pionic halo
 the dominant long-lived resonances are $\omega, \eta, \eta'$
and $K^0$, all with life-times $1/\Gamma_{res}$ greater than 20 fm/c).
If the emission function is a sum of the emission function of the
core and the halo,
\ben
	S(x; K) & = &
		S_{*,c}(\tau,\eta,r_x,r_y) + S_{h}(x; K),
\enn
and the Fourier-transformed emission function of the halo is sufficiently
narrow to vanish at the finite resolution of the  relative momentum $\Delta k$
in a given experiment, then one can show~\cite{halo} that
\ben
	N_1({\bf p}) & = &  N_{1,c}({\bf p}) +  N_{1,h}({\bf p}), \\
	C(\Delta k;K) & = & 1 + \lambda_*
	      {\dst \mid \tilde S_{*,c}(Q_\t, Q_\e, Q_\o, Q_\s) \mid^2 \ov
			\mid \tilde S_{*,c}(0,0,0,0) \mid^2 },
	\label{e:lbest}
\enn
where $N_{1,i}({\bf p})$ indicates the number of particles emitted
from the halo or from the core for $i = h,c$ and
the effective intercept parameter,
\ben
	\lambda_* & = & \lambda_*(K \simeq p) = \left[{\dst N_{1,c}({\bf p}) \ov
				 N_{1}({\bf p})} \right]^2
\enn
is the square of the ratio of the number of particles emitted from the core
to the number of all the emitted particles with a given momentum ${\bf p}$.
This effective intercept parameter arises due to the finite relative momentum
resolution and the comparably large region of homogeneity
characterizing the halo part of the system.

We would like to raise a warning flag here: The volume, which
the decay products of the long-lived resonances of a
given momentum are emitted from, is large only if the decaying
resonances have $\mid {\bf p} \mid / (m_{res} \Gamma_{res}) >>  1$ fm/c.
This in turn implies that the above simple picture
may need further corrections for  very low pt pions at rapidity $y = 0$.

There is a gap in the life-time distribution of abundant
hadronic resonances: $ 1 / \Gamma_{\rho } \simeq 1.3$ fm/c,
$1/\Gamma_{N*} \simeq 0.56$ fm/c, $ 1/\Gamma_{\Delta } \simeq 1.6 $ fm/c
and $ 1/\Gamma_{K*} \simeq 3.9 $ fm/c, which life-times are
of the same order of magnitude as the time-scales for re-scattering
at the time of the last hadronic interactions. These lifetimes are also all
a factor of 5 - 10 shorter than the life-time of the $\omega$ meson,
which is the long-lived resonance with shortest lifetime.
Thus the decay-product of the short-lived resonances  will mainly
contribute to the
core, which is resolvable by BEC measurements, while the decay-products
of long-lived hadronic resonances will mainly
 belong to the halo, re-defined alternatively as
the part of the emission function which is not resolvable
in a given  Bose-Einstein measurement.

\section{Classes of Simple Core Functions}
{\it If} the emission function of the core can be factorized,
\ben
	S_{*,c}(\tau,\eta,r_x,r_y) & = & H_*(\t) \, G_*(\e) \, I_*(r_x,r_y),
	\label{e:fact}
\enn
where $H_*(\t)$ stands for the effective emission function in proper-time,
$G_*(\e)$ stands for the effective emission function in space-time rapidity,
and $I_*(r_x,r_y)$ stands for the effective emission function in the
transverse directions, then the expression for the BECF can be further
simplified as
\ben
	C(\Delta k;K) = 1 +  \lambda_*\,
				{\dst \mid \tilde H_*(Q_\t) \mid^2 \,
				\mid \tilde G_*(Q_\e)  \mid^2 \,
				\mid \tilde I_*(Q_\o,Q_\s) \mid^2 \ov
				\mid \tilde H_*(0) \mid^2 \,
				\mid \tilde G_*(0)  \mid^2 \,
				\mid \tilde I_*(0,0) \mid^2 } .
		\label{e:lfact}
\enn
If the $I_*(r_x,r_y)$ function is symmetric for rotations in the
$(r_x,r_y)$ plane around its maximum point $r_{x,s}$ then one may introduce
$Q_t = \sqrt{Q_\s^2 + Q_\o^2} $ to find
\ben
	C(\Delta k;K) = 1 +  \lambda_*\,
				{\dst \mid \tilde H_*(Q_\t) \mid^2 \,
				\mid \tilde G_*(Q_\e)  \mid^2 \,
				\mid \tilde I_*(Q_t) \mid^2 \ov
			       \mid \tilde H_*(0) \mid^2 \,
				\mid \tilde G_*(0)  \mid^2 \,
				\mid \tilde I_*(0) \mid^2 }.
		\label{e:lfactt}
\enn
Such factorization around the saddle-point
happens e.g. for the new class of analytically solvable
models if certain
conditions are satisfied, as discussed in the subsequent part.
{}From the above expression it is clear that for this type of models
the dependence of the BECF on the components of the relative momentum
can be diagonalized with appropriate choice of the three independent
components of the relative momentum. Note that the assumed existence
of the Fourier-transformed distribution functions is a weaker condition than
the assumption of the analytic form of the Fourier-transformed function,
see Appendix for an example.
Another example was given e.g. in ref.~\cite{htau} for a $H(\tau)$ distribution
for which $\tilde H(Q_\t)$ is not analytic function at $Q_\t = 0$ and
$\mid H(Q_\t) \mid^2 $ does not start
with a quadratic term. In mathematical statistics
it is well known that
the Fourier-transformed stable distributions are not analytic at $Q = 0$
{}~\cite{lukacs}. On the other hand, there are many physically interesting
Gaussian models which correspond to the multivariate second order
Taylor expansion of the above general results, i.e.
the analytic form of the corresponding Fourier-transformed function.
The out-longitudinal cross term~\cite{uli} has been recently discovered also
in this context. To study the properties of the BECF let us apply a Gaussian
approximation to the effective distribution functions as
\ben
H_*(\tau ) & \propto & \exp(-(\t - \t_s)^2 / (2 \Delta \t_*^2) \, ),
	\label{e:hst} \\
G_*(\eta ) & \propto & \exp(-(\eta - \eta_s)^2 / (2 \Delta\e_*^2 \, ),
	\label{e:gst} \\
I_*(r_x,r_y) & \propto & \exp(-(\, (r_x - r_{x,s})^2 + (r_y - r_{y,s})^2 )
			/ (2 R_*^2) \, ). \label{e:ist}
\enn
Apart from the momentum-dependent parameters $\Delta \t_*, \Delta\e_*$
and $R_*$ the mean emission point may also be momentum-dependent in the
above expression, $\t_s = \t_s(K)$, $\eta_s = \eta_s(K)$,
$r_{x,s} = r_{x,s}(K)$ and $ r_{y,s} = r_{y,s}(K)$.
For the sake of simplicity we do not specify the normalization constants
in eq.~(\ref{e:ist}) since they cancel from the BECF which is given by
\ben
C(\Delta k;K) = 1 + \lambda_* \,
		\exp( - Q_\t^2 \Delta\tau_*^2 - Q_\e^2 \t_s^2 \Delta\e_*^2 -
		 Q_t^2 R_*^2). \label{e:c2di}
\enn
This is a diagonal form of BECF-s for which the factorization
property, eq.~(\ref{e:fact}) and the Gaussian approximation for the
core, eqs.~(\ref{e:hst}-\ref{e:ist}) are simultaneously satisfied.
In the present form of the BECF, there are no cross-terms among the chosen
variables. Now, let us rewrite this form using the standard
HBT coordinate system~\cite{bertsch} to find
\ben
	C(\Delta k;K) & = & 1 + \lambda_* \, \exp( - R_\s^2 Q_\s^2 -
			R_\o^2 Q_\o^2
		  - R_L^2 Q_L^2 - 2 R^2_{\o,L} Q_\o Q_L ), \label{e:crbecf} \\
	R_\s^2 & = & R_*^2, \label{e:rs} \\
	R_\o^2 & = &  R_*^2 + \delta R_\o^2, \label{e:ro} \\
	\delta R_\o^2 & = & \beta_t^2 ( \cosh^2[\e_s] \Delta\t_*^2 +
					\sinh^2[\e_s] \t_s^2 \Delta\e_*^2),
				\label{e:dro}\\
	R_L^2 & = &
		(\beta_L \sinh[\eta_s] - \cosh[\e_s])^2 \t_s^2 \Delta\e_*^2  +
			( \beta_L \cosh[\eta_s] - \sinh[\e_s])^2 \Delta\t_*^2,
			\label{e:rl}\\
	R_{\o,L}^2 & = & (\beta_t \cosh[\eta_s]
	( \beta_L \cosh[\eta_s] - \sinh[\e_s])) \Delta\t_*^2 + \nonumber\\
		\null & \null &                 (\beta_t \sinh[\eta_s]
		(\beta_L \sinh[\eta_s] - \cosh[\e_s]) ) \t_s^2 \Delta\e_*^2.
		\label{e:rol}
\enn
This result is non-perturbative in terms of the
variable $\eta_s$ and is valid in any frame.
The main limitations of this result are the assumed
Gaussian model-class, c.f. eqs.~(\ref{e:hst}-\ref{e:ist}) and
   the assumed smallness of
the emission region around $x_s(K)$ so that $t$ and $r_z$
dependencies could be linearized in terms of $\tau $ and $\eta$.

The above equations simplify a lot in the LCMS
system, where
$\beta_L = 0$:
\ben
	\delta R_\o^2 & = & \beta_t^2 ( \cosh^2[\e_s] \Delta\t_*^2 +
			\sinh^2[\e_s] \t_s^2 \Delta\e_*^2),  \label{e:ldro} \\
	R_L^2 & = & \cosh^2[\e_s] \t_s^2 \Delta\e_*^2 +
			\sinh^2[\e_s] \Delta\t_*^2, \label{e:lrl} \\
	 R_{\o,L}^2 & = & - \beta_t \sinh[\e_s] \cosh[\eta_s] (  \Delta\t_*^2 +
			 \t_s^2 \Delta\e_*^2). \label{e:lrol}
\enn
The life-time information
$\Delta\t_*^2$ and the invariant measure of the longitudinal size
along the $\t_s = const $ hyperbola, $\t_s^2 \Delta\e_*^2$
appear in a mixed form in the $ R_\o^2,  R_L^2$ and
the  $R_{\o,L}^2$ source parameters even in the LCMS frame.
The amount of these mixings is controlled by the value of $\eta_s^{LCMS}$.
This relationship clarifies the physical significance of the
$\eta_s^{LCMS}$, the space-time rapidity of the maximum of the emission
function in the LCMS frame: $\eta_s^{LCMS}$ is the {\it cross-term generating
hyperbolic mixing angle} for cylindrically symmetric, finite
systems undergoing longitudinal expansion and satisfying the
factorization property eq.~(\ref{e:fact}).
If $\eta_s^{LCMS} = 0$, no mixing of temporal and longitudinal components
appear in LCMS. In some limited sense one may call $\eta_s$ the cross-term
generating hyperbolic
 mixing angle in any frame, because if $\eta_s = 0$  in a certain
frame than cross-terms can be diagonalized away as follows.

Let us define the LSPS, the longitudinal saddle point system,
to be the frame where $\eta_s = 0$.
Since $\eta_s$ is a function of $K$
in a fixed frame,  $\eta_s = \eta_s(K)$, the LSPS
frame may depend on $K$ (e.g. on transverse mass of the pair).
 In the LSPS frame the out-long cross-term and the
mixing of the temporal and time-like informations can be diagonalized.
We have in LSPS
\ben
       \delta R_\o^2 & = & \beta_t^2 \Delta\t_*^2, \\
	 R_L^2 & = & \t_s^2 \Delta\e_*^2 + \beta_L^2 \Delta\tau^2_*, \\
	 R_{\o,L}^2 & = & \beta_t \beta_L \Delta\tau_*^2,
\enn
as follows from eqs.~(\ref{e:dro}-\ref{e:rol}).
Introducing the new variables $Q_0 = \beta_t Q_\o + \beta_L Q_L$
and $Q_t = \sqrt{ Q_\o^2 + Q_\s^2}$ we obtain for the correlation function
\ben
C(\Delta k;K) & = & 1 + \lambda_*
		\exp( - \Delta\tau^2_* Q_0^2 - \t_s^2 \Delta\e_*^2 Q_L^2 -
		R_*^2 Q_t^2).
		\label{e:clsps}
\enn
{}From this relationship we also see that
$Q_0 (LSPS) = Q_\t, Q_L(LSPS) = Q_\e$, c.f. eq.~(\ref{e:c2di}).

Let us study an expansion in terms of
 $\epsilon= \mid Y - y_0 \mid / \Delta \eta$,
 where $Y$ is the rapidity belonging
to $K$ the mean momentum of the pair,
and $\Delta \eta$ is the geometrical size of the expanding
system in the space-time rapidity variable,
satisfying $ \Delta\eta > \Delta\eta_* $.
It is obvious that in the LAB frame $\e_s^{LAB} =
 Y +{\cal O}(\epsilon)$, since in the
$\epsilon \rightarrow 0 $ limit we recover boost-invariance and the particle
emission must be centered around the only scale: the rapidity of the pair.
Similarly we see that $\e_s^{LCMS} = 0 + {\cal O}(\epsilon)$.
It follows that the cross-term and the crossing of temporal and
longitudinal information in the LAB frame is a leading order effect,
\ben
	\delta R_\o^2 & = & \beta_t^2 ( \cosh^2[Y] \Delta\t_*^2 +
					\sinh^2[Y] \t_s^2 \Delta\e_*^2)
			+ {\cal O}(\epsilon), \\
	R_L^2 & = & {\dst \t_s^2 \Delta\e_*^2 \ov \cosh^2[Y]}
			+ {\cal O}(\epsilon),  \\
	R_{\o,L}^2 & = & - \beta_t {\dst \sinh[Y] \ov \cosh[Y]}
			 \t_s^2 \Delta\e_*^2 + {\cal O}(\epsilon).
\enn
On the other hand, the mixing of the temporal and longitudinal
information is only next-to leading order in the LCMS
according to eq.~(\ref{e:lrol}),
i.e. $R_{\o,L}^2(LCMS)  =  0 + {\cal O}(\epsilon)$. However,
if the $\mid Y - y_0 \mid << \Delta \eta$ condition is not satisfied,
the out-long cross-term might be large even in LCMS, as has been
demonstrated numerically in ref.~\cite{chapman_95}.

The cross-term generating mixing angle
$\eta_s$ vanishes exactly in the LSPS frame,
becomes a small parameter in the LCMS if $\mid Y - y_0 \mid / \Delta \eta << 1$
and becomes leading order in any frame
significantly different from LSPS or LCMS.
Thus we confirm the recent finding ~\cite{LSPS},
that the out-longitudinal cross-term can be diagonalized away if
one finds the (transverse mass dependent) longitudinal rest frame
of the source.

Note, that cylindrical symmetry around the center of the particle emission
as assumed by eqs.~(\ref{e:hst}-\ref{e:ist})
is a stronger requirement than the cylindrical symmetry of the emission
function around the beam axis. This latter symmetry  implies only
that both the requirements $S(t,r_x,r_y,r_z;\, K_0,K_{out},K_L) = $
$S(t,\, -r_x,r_y,r_z;\, K_0, \, -K_{out},K_L)$ and
$S(t,r_x,r_y,r_z;\, K_0,K_{out},K_L) =$
$S(t,r_x,\, - r_y,r_z;\, K_0, K_{out},K_L)$ should be simultaneously
fulfilled. Thus cylindrical symmetry around the beam axis
is compatible with a different Gaussian radius in the
side and the out direction,
\ben
	I_*(r_x,r_y) \propto
		\exp\l( - {\dst (r_x - r_{x,s}(K) )^2 \ov 2 R_{*,x}^2}
		      - {\dst r_y^2 \ov 2 R_{*,y}^2 } \r),
		      \label{e:ista}
\enn
with $R_{*,x} \neq  R_{*,y}$. Cylindrical symmetry around
the beam axis implies only that $  r_{x,s}(K_0,K_{out},K_L)  =
-  r_{x,s}(K_0,\, - K_{out},K_L)$ and $r_{y,s}(K) = 0$.
In the low transverse momentum limit, when $K_{out} = 0$,
the relations  $r_{x,s}(K_{out} = 0) = 0$ and
$R_{*,x}(K_{out} = 0)  = R_{*,y}(K_{out} = 0)$ also follow
from cylindrical symmetry around the beam axis.
If $R_{*,x} \neq  R_{*,y}$ at a given non-vanishing
value of the mean transverse momentum $K_{out}$,
the generalized version of eqs.~(\ref{e:c2di},\ref{e:clsps})
for the BECF reads as
\ben
C(\Delta k;K) & = & 1 + \lambda_* \exp( - \Delta\tau_*^2 Q_{\tau}^2
					- \Delta\eta_*^2 \tau_s^2 Q_{\eta}^2
					- R_{*,x}^2 Q_{out}^2
					- R_{*,y}^2 Q_{side}^2 ).
\enn
In such a case, eqs. ~(\ref{e:rs},\ref{e:ro}) are also modified as
\ben
	R_{side}^2 & = & R_{*,y}^2, \\
	R_{out}^2  & = & R_{*,x}^2 + \delta R_{out}^2.
\enn
This implies that the difference between the {\it out} and {\it side}
radius parameters is {\it not  restricted }
by cylindrical symmetry around the beam axis to positive values only,
since $ R_{out}^2 -  R_{side}^2 =  \delta R_{out}^2 +  R_{*,x}^2 -  R_{*,y}^2$
which can also be negative
if $R_{*,y}$ is sufficiently large~\cite{wiedemann}.
However, cylindrical symmetry does imply that  $R_{out} =  R_{side}$ in the
$K_{out} \rightarrow 0 $ limit.

Up to this point, we have reviewed the
properties of BECF-s without reference to any particular
model, for some more and more limited classes of simple emission
functions. We have obtained certain model-independent relations
c.f. eqs.~(\ref{e:best},\ref{e:lbest},\ref{e:lfactt})
which are valid for some non-Gaussian as well a Gaussian source functions.
We have studied the relations between source parameters with a method
which is non-perturbative in terms of $\eta_s$, but perturbative
in terms of $\Delta\eta_*^2$ and $\Delta\tau_*^2 /\tau_s^2$.

Let us study the properties of an analytically solvable model-class
in the subsequent parts.

\section{A new class of analytically solvable models}
\label{s:newcl}
For central heavy ion collisions at high energies the beam or $r_z$ axis
becomes a symmetry axis. Since the initial state of the reaction is
axially symmetric and the equations of motion do not break this pattern,
the final state must be axially symmetric too.
However, in order to generate the thermal length-scales in  the transverse
directions, the flow-field must be either three-dimensional,
or the temperature distribution must have significant gradients in the
transverse directions. Furthermore, the local temperature
may either increase during the duration of the particle
emission because of the re-heating of the system caused by the
hadronization~\cite{zsenya} and/or intensive re-scattering processes
or decrease because of the expansion
and the emission of the most energetic particles from the
interaction region. An example for such a time-dependent
temperature was given e.g. by the solid line of Fig. 1. in ref.~\cite{TD}.

We study the following model emission function for high energy heavy ion
reactions:
\begin{eqnarray}
S(x;K) \, d^4 x & = & {\dst  g \ov (2 \pi)^3} \,  m_t \cosh(\eta - y) \,
\exp\l( - {\dst K \cdot u(x) \ov  T(x)} + {\dst \mu(x) \ov  T(x)}\r)
  \, H(\tau) d\tau  \, \tau_0 d\eta \, dr_x \, dr_y.
	\label{e:model}
\end{eqnarray}
Here $g$ is the degeneracy factor,
the pre-factor $m_t \cosh(\eta - y)$ corresponds to the flux of the
particles through a $\tau = const$ hypersurface according to the
Cooper-Fry formula and the four-velocity $u(x)$ is
\ben
u(x) & \simeq & \l( \cosh(\eta)
	\l(1 + b^2 \, {\dst r_x^2 + r_y^2 \ov 2 \tau_0^2}\r) ,
	\,\, b \,{\dst r_x \ov \tau_0}, \,\, b \,{\dst r_y \ov  \tau_0}, \,\,
	\sinh(\eta)
	\l( 1 + b^2 \,{\dst r_x^2 + r_y^2 \ov 2 \tau_0^2} \r) \, \r),
\enn
which describes a scaling longitudinal expansion with
a linear transverse flow  profile. The transverse flow is assumed
to be non-relativistic in the region where there is a significant
contribution to particle production.
The local temperature distribution $T(x)$ at the last interaction points
is assumed to have the form
\ben
{\dst 1 \ov T(x)} & = & {\dst 1 \ov  T_0 }
	\l( 1 + a^2 \, {\dst  r_x^2 + r_y^2 \ov 2 \tau_0^2} \r)
	   \, \l( 1 + d^2 \, {\dst (\tau - \tau_0)^2 \ov 2 \tau_0^2  } \r),
\enn
and the local rest density distribution is controlled by the
chemical potential $\mu(x)$ for which we have the ansatz
\ben
{\dst \mu(x) \ov T(x) }  & = &  {\dst \mu_0 \ov T_0} -
	{ \dst r_x^2 + r_y^2 \ov 2 R_G^2}
	-{ \dst (\eta - y_0)^2 \ov 2 \Delta \eta^2 }.
\end{eqnarray}
	The parameters $R_G$ and $\Delta\eta$ control the
	density distribution with finite geometrical sizes.
	The proper-time distribution of the last interaction points
	is assumed to have the following simple form:
\ben
	H(\tau) & = & {\dst 1 \ov ( 2 \pi \Delta \tau^2) ^{(1/2)} }
			\exp( - (\tau - \tau_0)^2 / (2 \Delta \tau^2) ) ).
\enn
	The parameter $\Delta\tau$ stands for the width of the
	freeze-out hypersurface distribution, i.e. the emission
	is from a layer of hypersurfaces which tends to
	an infinitely narrow hypersurface in the $\Delta\tau \rightarrow 0$
	limit.

	This emission function corresponds to
	a Boltzmann approximation to the local momentum distribution
	of a longitudinally expanding finite system which expands
	into the transverse directions with a transverse flow
	which is non-relativistic at the saddle-point.
	The transverse gradients of the
	local temperature at the last interaction points are controlled
	by the parameter $a$.
	The strength of the flow is controlled
	by the parameter $b$.
	The parameter $c = 1$ is reserved to denote the speed of light,
	and the parameter $d$ controls the strength of the change of the
	local temperature during the course of particle emission.

	Note that other shapes of the temperature profile lead to the
	same result if  $1/T(x)$ starts with the same second order
	Taylor expansion around $r_x = r_y = 0$. The physical significance
	of the transverse temperature profile is that it concentrates
	the emission of the particles with high transverse mass to
	a region which is centered around $r_x = r_y = 0$ and which
	narrows as the transverse mass increases. The Gaussian approximation
	to the inverse temperature profile is thus a technical
	simplification only, other decreasing temperature profiles
	have similar effects as follows from the above picture.
	Similarly, the significance of the temporal changes of the
	temperature is that it creates different effective emission times
	for particles with different transverse mass, the Gaussian
	approximation  is  technical simplification.

	For the case of $a = b = d = 0$ we recover the case of longitudinally
	expanding finite systems as presented in ref.~\cite{1d}.
	 The finite geometrical and temporal
	length-scales are represented by the transverse geometrical size
	$R_G$, the geometrical width of the space-time rapidity distribution
	$\Delta \eta$ and the mean duration of the particle
	emission $\Delta \tau$.
	Effects arising from the finite longitudinal size
	 were calculated analytically first in ref.~\cite{ave}
	 in certain limited regions of the phase-space.
	We assume here that the finite geometrical
	and temporal scales as well as the transverse radius and proper-time
	dependence of the inverse of the local temperature can be
	represented by the mean and the variance of the respective variables
	i.e. we apply a Gaussian approximation, corresponding to the
	forms listed above, in order to get analytically tractable results.
	We have first proposed the $a = 0, b = 1$ and $d = 0$ version
	of the present model, and elaborated also the $a = b = d =0$ model
	~\cite{1d} corresponding to  longitudinally expanding finite
	systems with a constant freeze-out temperature and no transverse flow.
	Soon the parameter $b$ has been introduced~\cite{uli} and it has
	been realized that the maximum of the emission function for a given
	mean    momentum $K$ has to be close to the beam axis,
	fulfilling $ r_{x,s} << \tau_0$, in order to get a transverse
	mass scaling law for the parameters of the Bose-Einstein
	correlation functions in certain limiting cases ~\cite{qm95}.
	In this region around the beam axis,
	however, the transverse flow is non-relativistic
	~\cite{uli} even for the case $b = 1$
	if this region is sufficiently small.
	Yu. Sinyukov and collaborators classified the various
	cases of ultra-relativistic transverse flows~\cite{hhm:te},
	~\cite{akkelin},  and introduced a parameter which controls
	the transverse temperature profile, corresponding to the
	$a \ne b = 0$ case. We have studied~\cite{qm95} the
	model-class $ a\ne 0$, $ b\ne 0$ , $d=0$
	which we extend here to the $d \ne 0$
	case too.

 The integrals of the emission function  are evaluated
using the saddle-point method ~\cite{sinyukov,sinyu93,uli}.
The saddle-point coincides with the maximum of the emission function,
parameterized by $(\tau_s,\e_s,r_{x,s},r_{y,s})$.
These coordinate values solve simultaneously the equations
\ben
	{\dst \partial S \ov \partial \t } & = &
	 {\dst \partial S \ov \partial \e } =
	 {\dst \partial S \ov \partial r_{x} } =
	 {\dst \partial S \ov \partial r_{y} } = 0.
\enn
These saddle-point equations are solved in the LCMS, the longitudinally
comoving system, for $\eta_s^{LCMS} <<1 $ and $r_{x,s} << \tau_0 $.
The approximations are self-consistent if
$ \mid Y - y_0 \mid << 1 + \Delta \eta^2 m_t /T_0 - \Delta \eta^2$ and
 $\beta_t << \tau_s^2 \Delta\eta_*^2 / ( b R_*^2)$ which
for the considered model can be simplified as
 $\beta_t
 = p_t / m_t << (a^2 + b^2) / b/\max(1,a,b) $.
The transverse flow is non-relativistic at the saddle-point if
 $\beta_t  << (a^2 + b^2) / b^2/\max(1,a,b) $.
We assume that $\Delta\tau < \tau_0$ so that the Fourier-integrals involving
$H(\tau)$ in the $0 \le \tau < \infty$ domain can be extended to
the $ -\infty < \t < \infty $ domain.
The radius parameters are evaluated here to the leading order
in $r_{x,s}/\tau_0$. Thus terms of
${\cal O}(r_{x,s}/\tau_0)$ are neglected,
however we keep all the higher-order
correction terms arising from the non-vanishing value of $\eta_s$
in the LCMS. We calculate both the radius parameters
and the invariant momentum distribution
in Gaussian saddle-point approximation.
We shall discuss the limitations of the saddle-point method after
presenting these results on BECF and IMD.

 For the model of eq.~(\ref{e:model}) the saddle point approximation for
the integrals leads to an effective
 emission function which
can be factorized similarly to eq.~(\ref{e:fact}).
Thus the radius parameters of the model  are expressible in
terms of the homogeneity lengths
$\Delta\eta_*, R_* $, $\Delta\tau_*$
and the position of the saddle point $\eta_s$
i.e. the cross-term generating
hyperbolic mixing angle.
The saddle-point in LCMS is given by $\tau_s = \t_0$, $\eta_s^{LCMS} =
(y_0 - Y) / (1 + \Delta\eta^2 (1 / \Delta \eta_T^2 -1) )$, $r_{x,s} =
\beta_t b R_*^2 / (\tau_0 \Delta\eta_T^2)$ and  $r_{y,s} = 0$.
Note that the space-time rapidity of the saddle-point $\eta_s^{LCMS}$
depends on the boost-invariant difference $y_0 - Y$ which can be evaluated
in any frame.
The radius parameters or {\it lengths of homogeneity} \cite{sinyukov,uli}
are given in the LCMS by
eqs.~(\ref{e:crbecf}-\ref{e:ro},\ref{e:ldro}-\ref{e:lrol}),
and we obtain
\ben
{\dst 1 \ov R_*^2 } & = &
		{\dst 1 \ov R_G^2} +
		{\dst 1 \ov R_T^2 } \cosh[\eta_s^{LCMS}],
			\label{e:rst}\\
 {\dst 1 \ov \Delta \eta_*^2} & = &  {\dst 1 \ov \Delta \eta^2 } +
		{\dst 1 \ov \Delta \eta_T^2} \cosh[\eta_s^{LCMS}] -
			{\dst 1 \ov \cosh^2[\eta_s^{LCMS}]},
			\label{e:est}\\
{\dst 1 \ov \Delta \tau_*^2 } & = &
		{\dst 1 \ov \Delta\tau^2} + {\dst 1 \ov \Delta\tau_T^2}
			\cosh[\eta_s^{LCMS}]. \label{e:dtst}
\enn
where the {\it thermal length-scales} are given by
\ben
 R_T^2 & = & { \displaystyle\strut \tau_0^2 \over
	\displaystyle\strut a^2 + b^2 }
	 { \displaystyle\strut T_0 \over \displaystyle\strut M_t},
		\label{e:rt} \\
 \Delta\eta_T^2 & = & {\dst T_0 \ov M_t}, \label{e:etat}\\
 \Delta\tau_T^2 & = & {\dst \tau_0^2 \ov d^2} {\dst T_0 \ov M_t}.
			\label{e:dtaut}
\enn
Here $M_t = \sqrt{K_0^2 - K_L^2}$
is the transverse mass belonging to the mean momentum $K$.
In the region of the Bose-Einstein enhancement, where the relative
momentum of the pair is small, $M_t$ satisfies
$M_t  = {\dst 1 \ov 2} (m_{t,1} + m_{t,2})
 (1  + {\cal O} ( y_1 - y_2)  +
{\cal O}((m_{t,1} - m_{t,2})/(m_{t,1} + m_{t,2}))  )$.
Please note the distinction between the subscripts for transverse
direction, indicated by $_t$, and the subscripts for the
'thermal' scales indicated by $_T$.
It is timely to emphasize at this point that the parameters
of the Bose-Einstein correlation function coincide with the
(rapidity and transverse mass dependent)
{\it lengths of homogeneity} ~\cite{sinyukov}
in the source, which physically can be identified with that region
in coordinate space where particles with a given momentum are emitted
from. The above relations indicate that these lengths of homogeneity
for simple thermal models can be basically obtained from two type of scales
in the framework of the saddle-point method.
These scales have different momentum - dependence and are referred to
as 'thermal' and 'geometrical' scales.

In contrast to the homogeneity lengths which can be defined even
without thermalization, the 'thermal scales' cannot be introduced
without at least approximate local thermalization. Thus the
thermal scales originate from the factor
$\exp(- p \cdot u(x)  /T(x))$. They measure that region in space-time
where thermal smearing can compensate the change of the local momentum
distribution
which in turn is induced by either the gradients of the flow-field
or the gradients of the temperature field.
 This is to be contrasted to the 'geometrical'
scales, which originate from the $\exp( \mu(x) /T(x) )$ factor which
controls the density distribution. The geometrical
 scales can be interpreted as the
regions in space-time where there is significant density to have particle
emission. Obviously for locally thermalized systems both the geometrical
and the thermal scales influence the regions of homogeneity and
the smaller scale will be the dominant one.
Since the four-momentum $p$ is explicit
in the factor $\exp(- p \cdot u(x) /T(x))$, and enters the 'geometrical'
scales only through the momentum dependence of the saddle-point,
the momentum dependencies for the 'thermal' and 'geometrical' scales
shall be in general different from each other. Note also that in the above
expression for $\Delta\eta_*$  a third type of scale is also present
in the term $- 1/\cosh^2[\eta_s]$, which stems from the
$m_t \cosh[\eta-y]$ Cooper-Fry pre-factor in eq.~(\ref{e:model}).
 Thus this term is related to
the shape of the freeze-out hyper-surface distribution
( which distribution tends to a single
 hyper-surface if  $\Delta \tau \rightarrow 0$).

The parameters of the BECF-s are dominated
by the smaller of the geometrical and the thermal scales
not only in the spatial directions but in the temporal direction too
 according to
eqs.~(\ref{e:rst}-\ref{e:dtaut}).
These analytic expressions show that
{\it even a complete measurement of the
parameters of the BECF as a function of the mean momentum $K$
may not be sufficient to determine uniquely
 the underlying phase-space distribution}~\cite{qm95,sinyukov,uli,1d,nr}.
We also can see that the LCMS frame approximately coincides
with the LSPS frame for pairs with $\mid y_0 - Y \mid << 1 + \Delta\eta^2
M_t/T_0  - \Delta\eta^2$
and the terms arising from the non-vanishing values of $\eta_s$ can be
neglected.
In this approximation, the cross-term generating hyperbolic mixing angle
 $\eta_s \approx 0$
thus we find the leading order LCMS result:
\ben
	C(\Delta k;K) & = & 1 + \lambda_* \exp( - R_L^2 Q_L^2 - R_{\s}^2 Q_{\s}^2
			  - R_{\o}^2 Q_{\o}^2) ,
\enn
	with a vanishing out-long cross-term, $R_{\o,L} = 0$.
	To leading order,  the parameters of the correlation function
	are given by
\ben
	R_{\s}^2 & = & R_*^2, \label{e:side} \\
	R_{\o}^2 & = & R_*^2 + \beta_t^2 \Dt_*^2 ,\label{e:out} \\
	R_L^2 & = & \tau_0^2 \Delta\eta_*^2. \label{e:long}
\enn
	Observe that the difference of the side and the out radius
	parameters is dominated by the lifetime-{\it parameter} $\Delta\tau_*$.
	Thus a vanishing difference between the $R_\o^2 $ and $R_\s^2$
	can be generated dynamically if the duration
	of the particle emission is large, but the thermal duration
	$\Delta\tau_T$ becomes sufficiently small, c.f. eq.~(\ref{e:dtst}).
	 This in turn can be associated
	with intensive changes in the local temperature distribution
	during the course of the particle emission.

	Observe, that the BECF in an arbitrary frame can be
	obtained from combining eqs.~(\ref{e:rst}-\ref{e:dtaut})
	with the general expressions given by
	eqs.~(\ref{e:crbecf}-\ref{e:rol}).
	In that case, the value of $\eta_s = Y + \eta_s^{LCMS} =
	Y + (y_0 - Y) / (1 + \Delta\eta^2 (1 / \Delta \eta_T^2 -1) )$
	has to be used in eqs. ~~(\ref{e:crbecf}-\ref{e:rol}).

	Note also that in our results
	 higher order terms arising from the non-vanishing
	value of $\eta_s$ in the LCMS are summed up, while in refs.~\cite{uli}
	the first sub-leading corrections were found.

\section{Invariant momentum distributions} \label{s:imd}
The IMD plays a {\it complementary role}
to the measured Bose-Einstein correlation function ~\cite{1d,nr,qm95}.
Thus a {\it simultaneous analysis} of the Bose-Einstein correlation functions
and the IMD may reveal information
both on  the temperature and flow profiles  and on the geometrical sizes.

For the considered model, the invariant momentum distribution
can be calculated as
\ben
		N_{1,c}({\bf p}) & = &
	{\dst g \ov (2 \pi)^3 } \,\, \exp\l( {\dst \mu_0 \ov T_0} \r) \, \,
	m_t \,\, (2 \pi {\overline{\Delta\eta}_*}^2 \tau_0^2)^{1/2} \, \,
	(2 \pi R_*^2) \,\, {\dst \Dt_* \ov \Dt}
	 \cosh(\overline{\eta_s}) \, \,
	\exp(+ {\overline{\Delta\eta}_*}^2 / 2) \times  \nonumber \\
\null & \null & \times \exp\l( - {\dst (y - y_0)^2 \ov 2 (\Delta\eta^2 +
	\Delta \eta_T^2) } \r) \,
\exp\l( - {\dst m_t \ov T_0} \l(  1 - f \, {\dst \beta_t^2 \ov 2} \r) \r) \,
\exp\l( - f \, {\dst m_t \beta_t^2 \ov 2 (T_0+ T_G)} \r),
\label{e:imd}
\enn
where the geometrical contribution to the effective temperature is given by
$T_G(m_t)  = T_0 \,  R_G^2 / R_T^2(m_t) =
(a^2 + b^2)\, M_t \, R_G^2 / \tau_0^2$ and the fraction $f$ is defined as
$f = b^2 / (a^2 + b^2)$, satisfying $0 \le f \le 1$.
In  eq.~(\ref{e:imd})  the  quantities ${\overline{\Delta\eta}_*}^2$ and
$\overline{\eta_s}$
are defined as
\ben
{\dst 1 \ov {\overline{\Delta\eta}_*}^2 } & = &
	{\dst 1 \ov \Delta\eta^2} +
	{\dst 1 \ov \Delta\eta_T^2} \cosh[\overline{\eta_s}], \\
\overline{\eta_s} & = & (y_0 - y) / (1 + \Delta\eta^2  / \Delta \eta_T^2 ),
\enn
i.e. these quantities differ from the $\Delta\eta$ and $\eta_s^{LCMS}$
 by the contributions of the Cooper-Fry pre-factor. This happens because
eq.~(\ref{e:imd}) is obtained by applying the saddle-point method for
eq.~(\ref{e:imdd}) with the model emission function
eq.~(\ref{e:model}) in such a way that the Cooper-Fry pre-factor
$m_t \cosh[\eta - y]$ is kept exactly and the remaining factors
are approximated with the saddle-point technique in LCMS~\cite{qm95},
described in details  in the previous section.
Since the saddle-point equations are solved in LCMS in a region where
$\eta_s - y << 1$, a term in the exponent is approximated by
$ \cosh[\eta_s - y] \simeq 1 + (\eta_s - y)^2 / 2$.

For the considered
model, the rapidity-width
$\Delta y(m_t)$ of the invariant momentum
distribution at a given $m_t$,
shall be dominated by the {\it longer} of the thermal and geometrical
length-scales.
If the condition $a \simeq 0$ is fulfilled, i.e. $ f \approx 1$,
the longer of the thermal and geometrical scales shall also
dominate $ T_*$, the effective temperature (slope parameter)
at a mid-rapidity $y_0$. E.g. the following relations hold:
\begin{eqnarray}
\Delta y^2(m_t) & = & \Delta \eta^2 + \Delta \eta_T^2(m_t) \qquad
\mbox{\rm and} \qquad
{\dst 1 \ov T_*}  =  {\dst f \ov T_0 + T_G(m_t=m) } + {\dst 1 -f \ov T_0}.
	\label{e:tempe}
\end{eqnarray}
That is why the IMD measurements
can be considered to be complementary to the BECF data.

In the special limiting case when gradients of the temperature are negligible,
$a = 0$ and $f = 1$, we have $T_* = T_0 + m  b^2 R_G^2 / \tau_0^2$.
If the flow velocity at the
geometrical radius $\langle u_t \rangle \equiv b R_G / \tau_0 $
is independent of the particle type,
we obtain a relation
\ben
T_* & =  & T_0 + m \langle u_t \rangle^2.
	\label{e:us}
\enn
A similar relation for the mass dependence of the
mean (transverse) kinetic
energy was introduced in ref.~\cite{schneder}
for longitudinally expanding systems
with non-relativistic transverse flows.
This simple relation (~\ref{e:us})  for the effective temperature can
be considered as a special case of the more general eq.~(\ref{e:tempe}).

The measured IMD can be obtained from the IMD of the core as given
above and from the measured $\lambda_*({\bf p})$ parameters of the BECF as
\ben
	N_1({\bf p}) = {\dst 1 \ov \sqrt{\lambda_*({\bf p})} }
			 N_{1,c}({\bf p}),
\enn
see e.g. ref.~(\cite{halo}) for further details.

The invariant momentum distribution described by eq.~(\ref{e:imd})
features two types of low transverse momentum enhancement as
compared to  a static thermal source with a slope parameter $T_*$.
One may introduce the {\it volume factor} or $V_* (y,m_t)$
which yields the momentum-dependent size of the region,
where the particles with a given momentum are emitted from:
\ben
V_* (y,m_t) & = &
	(2 \pi {\overline{\Delta\eta}_*}^2 \tau_0^2)^{1/2} (2 \pi R_*^2)
		{\dst \Delta\tau_* \ov \Delta\tau}.
\enn
The {\it rapidity-independent low-pt enhancement} is a consequence
of the transverse mass dependence of this effective volume,
which
may depend on $m_t$ for certain limiting cases in the following ways
\ben
V_* (y,m_t) \propto \l({\dst T_0 \ov m_t} \r)^{k/2},
\enn
where $k = 0 $ for a static fireball ($ a=b=d=0 $  and $\Delta\e_T^2 >> \Delta
\e^2$).
The case $k = 1$ is satisfied for  $ a=b=d=0 $ and $\Delta\e^2  >> \Delta
\e_T^2$,  which describes long, one dimensionally
expanding finite systems~\cite{1d}.
The case $k = 2$ corresponds to
$ a = b = 0 \ne d$ and  $\Delta\e^2  >> \Delta\e_T^2 $,
describing longitudinally expanding systems with cooling.
The case $k = 3$ corresponds to $ a \ne 0$ , $b \ne 0 = d $ i.e.
three-dimensionally expanding,
cylindrically symmetric, finite systems
possibly with a transverse temperature profile
{}~\cite{qm95} and the $k = 4$ case corresponds to the same but with
a $d \ne 0 $ parameter, describing the temporal changes in the
local temperature during the particle emission process appended with the
condition $\Delta\t_T << \Delta\tau$.
Thus the inclusion of this effective $m_t$ dependent volume factor
into the data analysis not only would undoubtedly increase the
precision of the measurements of the slope parameters,
but in turn it also could shed light on the
dynamics of the particle emission from such complex systems.

The {\it rapidity-dependent low-pt enhancement}, which is a generic
property of the longitudinally expanding finite systems~\cite{lpte},
reveals itself in the rapidity-dependence of the effective temperature,
defined as the slope of the exponential factors in the IMD
 in the low-pt limit at a given value of the rapidity.
The leading order ~\cite{lpte} result is
\ben
T_{eff}(y) & = & {\dst T_* \ov 1 + a (y - y_0)^2 }
\qquad \mbox{\rm with} \qquad a = {\dst T_0 T_* \ov 2 m^2}
	\l(\Delta\eta^2 + {\dst T_0 \ov m}\r)^{-2}.
\enn
Please note that this analysis of the low transverse mass region
of the IMD relies on the
applicability of the saddle-point method
in the low transverse momentum
region too. Thus it may be valid for kaons or heavier particles
(as well as locally very cold pionic systems).
However, in case of pions, the self-consistency
of the applied formulas and their region of validity
 has to be very carefully checked.
This region of the applicability of the
 saddle-point technique for the considered
finite systems is discussed in detail in the next section.
Note that the low transverse momentum region is populated by
a number of resonance decays. For the long-lived resonances,
thus a non-trivial $1/\sqrt{\lambda_*({\bf p})}$ factor may appear and
contribute to both the rapidity-dependent and the rapidity-independent
low-$p_t$ enhancement. Although this factor is measurable
from the shape-analysis of the BECF,
care is required to study the contribution of the decay products
of short lived resonances to the momentum distribution  of pions.
Kaons or other heavier particles thus provide a cleaner test for these
analytic results as compared to pions.

The {\it high-pt enhancement or decrease}
refers to the change
of the
effective temperature at mid-rapidity with increasing $m_t$ .
The large transverse mass limit
$T_{\infty}$  shall be in general
different from the effective temperature at low pt given by $T_*$
since
\ben
T_{\infty} & = & {\dst 2 T_0 \ov 2 - f} \qquad \mbox{\rm and}\qquad
{\dst T_{\infty} \ov T_* }  =  {\dst 2 \ov 2 - f} \l( 1 - f {\dst T_G(m)
	\ov T_0 + T_G(m) }\r).
\enn
Utilizing $T_G / T_0 = R_G^2 / R_T^2$, the
high-pt enhancement or decrease turns out to be controlled by the
ratio of the thermal radius $R_T(m_t = m)$ to the geometrical radius $R_G$.
One obtains $T_{\infty} > T_* $ if $ R_T^2(m) > R_G^2$
and similarly $T_{\infty} < T_* $ if $ R_T^2(m) < R_G^2$.
Since for large colliding nuclei $R_G$ is expected to increase,
a possible high-pt decrease in these reactions may become a geometrical
effect, a consequence of the large size.

\section{Limitations}
The simple analytic formulas presented in the previous sections
are obtained in
a saddle-point approximation for the evaluation of the space-time
integrals. This approximation is known to converge to the
exact result in the limit the integrated function develops a
sufficiently narrow peak, i. e. both
$\Delta\eta_*^2 << 1$ and $\Delta \tau_*^2 / \tau_s^2 << 1$
are required.
This in turn gives a lower limit in $m_t$ for the applicability
of the formulas for the class of models presented in the previous section.

[However, the saddle-point method may give precise results even
when the integrand does not develop a narrow peak.
For the emission function of ref.~\cite{nr} the saddle-point method
gives exact result, because that  emission function can be re-written
in Gaussian form.]

{}From the requirement $\Delta\eta_*^2 << 1$ we have
\ben
	m_t \cosh[\eta_s]/ T_0  & >> &  1 + 1/\cosh^2[\eta_s] - 1/\Delta\eta^2.
	\label{e:mtceta}
\enn
{}From the requirement $\Delta \tau_*^2 / \tau_s^2 << 1$ one gets
\ben
	 m_t \cosh[\eta_s] / T_0 & >> & ( 1 - \tau_s^2/\Delta\tau^2) / d^2.
	\label{e:mtctau}
\enn
These are the conditions governing the validity of
the calculation of the parameters of the BECF as presented in
section~\ref{s:newcl}. Compared to the condition ~(\ref{e:mtceta}), the
condition of validity of the
calculation of the invariant momentum distribution is less
stringent, since one needs to satisfy only $\overline{\Delta\eta}_*^2 << 1$,
which yields
\ben
	m_t \cosh[\overline{\eta}_s]/ T_0  & >> &  1  - 1/\Delta\eta^2.
	\label{e:mtcetabar}
\enn

For the case of the NA44 measurement one may {\it estimate} the
region of reliability for the analytic formulas presented in the previous
parts using inequalities eq.~(\ref{e:mtceta},\ref{e:mtctau}).
Since the data indicate $R_{side} \simeq R_{out}$ within errors,
the inequality $\Delta \tau_*^2 / \tau_s^2 << 1$ and its consequence,
eq. ~(\ref{e:mtctau}) seems to be well justified. In the inequality
eq. ~(\ref{e:mtceta}) the finite longitudinal size plays an important
role. For infinite systems, $\Delta \eta = \infty$, the calculations
of BECF parameters are reliable for $ m_t >> 2 T_0$ (since $\eta_s = 0$ for
infinite systems
in LCMS), while the calculations for the IMD are
reliable to  $ m_t >> T_0$.
In the mid-rapidity region where NA44 data were taken,
one has $\eta_s \simeq 0 $ and for {\it finite} systems
one finds $ m_t >> T_0 (2 - 1 /\Delta\eta^2)$.
Note that this estimated lower limit in $m_t$ is extremely sensitive to
the precise value of $\Delta\eta$ in the region $\Delta\eta \simeq 1 /\sqrt{2}
\approx 0.7$.
Thus for finite systems the region of applicability of our results
extends to lower values of $m_t$ than for infinite systems which were
recently studied in great detail in ref.~\cite{wiedemann}.
The inequality eq.~(\ref{e:mtceta})
 can be used in basically two ways: either one assumes
a value for $T_0$ and then obtains the lower limit in $m_t$
for the applicability of the saddle-point method, or one assumes
that the saddle-point method is applicable to a certain value of $m_t$
(e.g. in case it gives a good description of data),
and then one obtains an upper limit for the corresponding $T_0$ parameter,
the central temperature at the mean time of last interactions.

An upper transverse momentum limit
is obtained for the validity of the calculations
in sections ~\ref{s:newcl}-\ref{s:imd} from the requirement
$r_{x,s} / \tau_0 < 1$ or  $r_{x,s}^2 / \tau_0^2 << 1$.
This condition and the requirement $ b r_{x,s} / \tau_0 < 1$
are fulfilled simultaneously if
\ben
	\beta_t << {\dst a^2 + b^2 \ov b \max(1,b) \max(1,a,b)}.
\enn
If $a^2 + b^2 \simeq 1$ this condition simplifies to
$\beta_t << 1/b $.

When comparing to data, detailed numerical studies may be
necessary~\cite{wiedemann} to check the precision of the saddle-point
integration. In a subsequent paper we plan to present these studies
together with a detailed comparison of the model to the available NA44 data.

\section{Limiting cases}

	Observe that both the thermal and the
	geometrical length-scales enter both the   parameters
	of the Bose-Einstein correlation
	function and those of the invariant momentum
	distribution.
	We limit the discussion in this part to the mid-rapidity region
	$\eta_s^{LCMS} \simeq 0$ and we assume $\Delta\eta_*^2 << 1$,
	i.e. we neglect the $1/\cosh^2[\eta_s]$ term in eq. ~(\ref{e:est}).
	Various limiting cases can be obtained as combinations of basically
	the relative size of the thermal and the geometrical scales
	in the transverse, longitudinal and temporal directions.
	These in turn are:

	{\it i)} If $R_T(M_t)  >> R_G$ in a certain $M_t$ interval,
	we have also $ T_0 >> T_G(m_t) $ at the same
	transverse mass scale. In this region, the side radius parameter
	shall be determined by the geometrical size
	$R_{\s} = R_* \simeq R_G$, hence it shall be transverse mass
	independent.

	The $m_t$ distribution at mid-rapidity shall be proportional to
	$\exp(-m_t / T_0)$.

	{\it ii)} If $\Delta\eta_T >> \Delta \eta$,
	we have $R_L \simeq \tau_0 \Delta\eta$ and the rapidity-width of the
	IMD shall be dominated by the thermal scale,
	$\Delta y^2(m_t) \simeq \Delta\eta_T^2 = T_0/m_t$.

	{\it iii)} If $\Delta\tau_T >> \Delta \tau $,
	the temporal duration shall be measured by
	$ R_{\o}^2 - R_{\s}^2 \simeq \beta_t^2 \Delta\t^2$.
	The invariant momentum distribution shall be influenced only
	through the $\Delta\t_*  / \Delta\t \approx 1$ factor
	in $V_*$.

	These cases are rather conventional limiting cases.
	An unconventional limit complements each:

	{\it iv)}  If $R_T(M_t)  << R_G$ in a certain $M_t$ interval,
	we have also $ T_0 << T_G(m_t) $ at the same
	transverse mass scale. In this region, the side radius parameter
	shall be determined by the thermal size
	$R_{\s} = R_* \simeq R_T(M_t)$,
	hence it shall be transverse mass dependent,
	 $R^2_{side} \propto \tau_0^2 T_0 /M_t$.

	The $m_t$ distribution at mid-rapidity shall be proportional to
	$\exp(-m_t / T_*)$. If $a^2 << b^2$, we have $T_* \simeq T_G(m)$
	as follows from eq.~(\ref{e:tempe}).

	{\it v)}  If $\Delta\eta_T << \Delta \eta$,
	we have the leading order LCMS result
	$R_L \simeq  \tau_0 \Delta\eta_T \simeq
	 \tau_0 \sqrt{T_0 /M_t} $, and the
	 rapidity-width of the IMD shall be dominated by the geometrical scale,
	$\Delta y^2\simeq \Delta \eta^2$.

	{\it vi)}  If $\Delta\tau_T << \Delta \tau $,
	the {\it thermal} duration shall be measured by
	$ R_{\o}^2 - R_{\s}^2 \simeq \beta_t^2 \Delta\t^2_T  \simeq \beta_t^2
\tau_0^2 T_0
	/ ( d^2 M_t)$. For large values of the transverse mass, the
model thus shall feature a dynamically generated vanishing
duration parameter, which has a specific transverse mass dependence.
	The invariant momentum distribution shall be influenced only
	through the $\Delta\t_*  / \Delta\t \simeq 1/\sqrt{m_t}$ factor
	in $V_*$.

	Some combinations of cases {\it i) -- vi),}
	are especially interesting, as:

	{\it vii)} If all the finite geometrical source sizes,
	$R_G, \Delta \eta$ and $\Delta \tau$
	 are large compared to the corresponding thermal
	length-scales we have in LCMS
\ben
	\Dt_*^2 &  \simeq & {\dst \tau_0^2 \ov d^2} \, {\dst T_0\ov M_t}, \\
	R_L^2 & \simeq & {\tau_0^2 } \, {\dst T_0\ov M_t},\\
	R_\s^2 & \simeq & {\dst \tau_0^2 \ov a^2 + b^2 } \, {\dst T_0\ov M_t}.
\enn
	Thus if $d^2 >> a^2 + b^2 \approx 1$ the model may  feature
	{\it a dynamically generated vanishing duration parameter}.
	In this case, the model
	{\it predicts an $M_t$ - scaling for the duration parameter}
	as
\ben
	\Delta\t_*^2 & \propto  & {\dst 1 \ov
			{M_t}}.
\enn
	This prediction could be checked experimentally
	if the  error bars of the
	measured radius parameters were decreased to such a level that
	the difference between the out and the side radius parameters
	would be significant.

	Alternatively, if the vanishing duration parameter of the BECF
	is generated due to a very fast hadronization process as discussed
	in ref.~\cite{TD}, then one has
\ben
	 \Delta\t_*^2 \simeq \Delta\t^2 \propto  const,
\enn
	i.e. in this case the duration parameter becomes independent
	of the transverse mass.

	If the finite source sizes are large compared to the thermal
length-scales and if we also have $a^2 + b^2 \approx 1$,
one obtains an
$M_t$ -{\it scaling} for the parameters of the BECF,
\ben
R_{side}^2 &\simeq & R_{out}^2 \simeq R_L^2 \simeq \tau_0^2 {\dst T_0 \ov M_t},
\quad \mbox{\rm valid for} \quad \beta_t << {\dst 1 \ov b}.
\label{e:mtsc}
\enn

	Note that this relation is
independent of the particle type and has been observed
in the recent NA44 data in $ S+Pb$ reactions at CERN SPS~\cite{na44}.
Preliminary NA49 data for $Pb + Pb$ at CERN SPS are also compatible
with this scaling law~\cite{na49}.
This $M_t$-scaling may be valid to arbitrarily large
transverse masses with $\beta_t \approx 1$ if $b << 1$.
The lower limit of the validity of this relation is given by the
applicability of the saddle-point method,
eqs.~(\ref{e:mtceta},\ref{e:mtctau}).
To generate  a vanishing difference between the
side and out radius and an $M_t$-scaling for the BECF radii
simultaneously,
the parameters have to satisfy the inequalities
{}~(\ref{e:mtceta},\ref{e:mtctau}) as well as
	$ b << a^2 + b^2 \approx 1 << d^2$,
i.e. the cooling should be the fastest process, the next dominant
	process within this phenomenological picture has to be the development
	of the transverse temperature profile and
	finally the transverse flow shall be relatively weak.
	If the temporal changes of the temperature are not intensive enough
	than a small life-time parameter can
	also be obtained by a fast hadronization and
	simultaneous freeze-out as discussed in
	ref.~\cite{TD} with $\Delta \tau \approx 0$.

	We would like to emphasize that there are
	{\it a number of conditions} in the model
	which need to be satisfied simultaneously to get the scaling behavior,
	which is supported by 9 NA44 data-points (3 for kaons
	and 6 for pions, ref.~\cite{na44}).
	One has to wait for {\it future data points} to learn more
	about the experimental status of the scaling. The model presented in
	this paper {\it may describe more complex}
	 transverse momentum dependences
	of the parameters of the Bose-Einstein correlation function, too,
	the $M_t$-scaling is only one of its virtues
	in a specific limiting case.
	However, it is rather difficult to get a limiting case with
	$R_L \approx R_{side} \approx R_{out} \propto 1 /\sqrt{M_t}$
	in analytically solvable models. Such a behavior is related to
	the cylindrical symmetry of the emission function.

	Thus the symmetry of the BECF in LCMS can be considered as
	 a strong indication
	for a
	{\it three-dimensionally expanding, cylindrically symmetric}
	source, possibly with a transverse and temporal temperature
	profile. The LCMS frame is selected if the mean emission point
	or saddle-point stays close to the
	symmetry axis even for particles with a large transverse mass,
	 $r_{x,s}(m_t) << \tau_0$ and if the finite longitudinal
	size introduces only small difference between the LCMS and LSPS
	frames, i.e. $\mid y - y_0 \mid << 1 + \Delta\eta^2( m_t / T_0 - 1)$ .
	In the case considered in section ~\ref{s:newcl}
	the emission function
	is cylindrically symmetric and so the BECF is symmetric in the LCMS
	of the pair ( and {\it not} in the center of mass system of the pair
	~\cite{glasgow}).

	{\it viii)}
	It is interesting to investigate the other limiting case when
	$R_T >> R_G, \Delta\eta_T >> \Delta\eta$ and $\Delta\tau_T >> \Dt$
	by combining the limiting cases {\it i, - iii}.
	In this case one obtains
\ben
	R_L^2  \simeq \t_0^2 \De^2 = R_{L,G}^2, \quad\quad\quad &
	R_{\s}^2  \simeq R_G^2,& \quad\quad\quad
	R_{\o}^2  \simeq  R_G^2 + \beta_t^2 \Dt^2,\\
	\Dy^2 (m_t)  \simeq  \Det^2 = {\dst T_0 \ov m_t}, &
		 \quad\quad\quad & T_*  =  T_0.
\enn
	Thus, if the thermal length scales are
	larger than the geometrical
	sizes in all directions, the BECF measurement determines
	the geometrical sizes properly, and the $p_t$ and the $dn/dy$
	distributions are determined by the temperature of the source.
	In this case the momentum distribution reads as
\ben
	N_1({\bf p}) \propto m_t \cosh( y - y_0) \,
		\exp\l(- {\dst m_t \cosh( y - y_0)\ov T_0} \r),
\enn
	which is a thermal distribution for a static source
	located at the mid-rapidity $y_0$.

	Thus {\it two length-scales are present}
	in all the three principal directions of
	three-dimensionally expanding systems.
	The BECF radius parameters are
	dominated by the {\it shorter} of the thermal
	and geometrical length scales.
	However, the rapidity-width of the $d^2 N / dy /dm_t^2$
	distribution,
	$\Delta y^2(m_t)$ is
	the quadratic sum of the geometrical and the thermal
	length scales, thus it is dominated by the {\it longer} of the two.
	Similarly, the effective temperature is
	dominated by the {\it higher} of the two
	temperature scales for $f \approx 1$ according to
	eq.~(\ref{e:tempe}). The effective temperature of the $m_t$
	distribution is  decreasing
	in the target and projectile rapidity region in this class
	of analytically solvable models.

	This study is a generalization of
	the basic ideas presented and illustrated in ref.~\cite{nr}
	for the case of three-dimensionally expanding, cylindrically
	expanding finite systems with a scaling longitudinal flow,
	weak transverse flow and a transverse and temporal temperature
	profile.

\section{Summary}
A general
formulation is presented for the two-particle Bose-Einstein correlation
function for cylindrically symmetric systems undergoing
collective hydrodynamic expansion, c.f. eqs.
{}~(\ref{e:best},\ref{e:lbest},\ref{e:lfact},\ref{e:lfactt}).
Note that these relations were shown to be
valid for certain limited classes of emission functions. The resulting
class of Bose-Einstein correlation functions,
however, includes non-Gaussian correlation functions too.
The case of Gaussian correlation functions is studied in detail
and the radius parameters are expressed in the LAB, LCMS and LSPS systems,
where the functional form of the correlation functions
becomes more and more simplified.
The cross-term generating hyperbolic mixing angle is identified
with the value of the  $\e$ variable of the saddle point
in the considered frame.

	A class of Gaussian
	models is introduced which in some regions of the model-parameters
	may obey an {\it $M_t$-scaling for the side, out and longitudinal
	radius parameters}.
		Vanishing effective duration of the particle emission
	may be generated by the temporal changes of the local temperature
	during the evaporation. The model predicts an {\it $M_t$-scaling
	also for the duration parameter} in this limiting case.

	Finally we stress that {\it both} the invariant momentum distribution
	and the Bose-Einstein correlation function may carry only
	partial information about the phase-space distribution
	of particle emission. However, their {\it simultaneous
	analysis} sheds more light on the dynamics and
	the geometrical source-sizes.
\medskip\medskip\bigskip
\begin{center}
{\bf Acknowledgments}
\end{center}
\medskip
Cs. T. would like to thank M. Asakawa, G. Gustafson,
M. Gyulassy, U. Heinz, S. Hegyi, Y. Pang, Yu. Sinyukov
and A. Wiedemann for stimulating discussions, B. L\"orstad
and M. Gyulassy for
hospitality at Lund and Columbia University.
This work was supported
by the Human Capital and Mobility (COST) program of the
EEC under grant No. CIPA - CT - 92 - 0418
(DG 12 HSMU), by the Hungarian
NSF  under Grants  No. OTKA-F4019, OTKA-T2973 and
OTKA-W01015107
and by the Hungarian - US Joint Fund MAKA 378/93.

\medskip
\medskip
\begin{center}
{\bf APPENDIX}
\end{center}
\medskip
In this appendix we give a simple example when
the Fourier-transformed emission function exists but the
Gaussian version of the saddle-point method is not applicable.
Let us consider the one-dimensional Lorentzian distribution function
\ben
S(r) & = & {\dst 1 \ov \pi R } \,\, {\dst 1 \ov (1 + r^2 / R^2)}.
\enn
Here $r$ is a real variable (in one dimension).
The corresponding correlation function is
\ben
C(q) & = & 1 + \mid \tilde S(q) \mid^2,
\enn
with
\ben
\tilde S(q) & = & \int_{-\infty}^{\infty} dr \, S(r) \exp(- i q r),
\enn
which yields
\ben
C(q) & = & 1 + \exp( - 2 \mid q \mid R).
\label{e:cexp}
\enn
This function is not analytic at $q = 0$ because it depends on the
modulus of $q$, and for positive values of $q$ its Taylor expansion
starts with a linear term.
This is to be contrasted with the results for the saddle-point method.
If the Gaussian version of the saddle-point  method is applicable, then $\tilde
S(q)$ can be expanded
into
a Taylor series around $q = 0$ as
\ben
\tilde S(q) & =& 1 + i \langle r \rangle q - \langle r^2 \rangle q^2 /2 + ...
	\,\, .
\enn
Here the average of a function of variable $r$  is defined as
\ben
 \langle f(r) \rangle & = &  \int_{-\infty}^{\infty} dr \, f(r) S(r)
\enn
and the two-particle correlation function can be written as
\ben
C(q) &  \approx & 2 - q^2 R_G^2 \approx 1 + \exp(- q^2 R_G^2)
\enn
with
\ben
R^2_G & = & \langle r^2 \rangle -  \langle r \rangle^2.
\enn
Since for the considered function $R^2_G = \infty$,
the Gaussian saddle--point method is not applicable.
Still, the Fourier-transformed emission function and the
BECF exist, as given by eq.~(\ref{e:cexp}).

Cases similar to this are characterized by
non-Gaussian correlation functions~\cite{lukacs}.
Similar examples can be found among multi-variate
distributions, too.

\vfill
\eject
\begin{thebibliography}{99}
\referencestyle
\bibitem{HBT}            R. Hanbury-Brown and R. Q. Twiss, Phyl. Mag.
	{\bf 45} 663 (1954); R. Hanbury-Brown and R. Q. Twiss, Nature
	(London) {\bf 177}, (1956) 27 and {\bf 178} (1956) 1046
\bibitem{QM}            Proceedings of the Quark Matter conferences,
			especially Nucl.\ Phys.\  {\bf A498},
			(1989), Nucl.\ Phys.\ {\bf A525}, (1991),
			Nucl.\ Phys.\ {\bf A544}, (1992)
			Nucl.\ Phys.\ {\bf A566}, (1993) and Nucl. Phys.
			{\bf A590} (1995)
\bibitem{na44}          H. Beker et al, NA44 Collaboration,
			Phys. Rev. Lett. {\bf 74} (1995) 3340
\bibitem{bengt}         B. L\"orstad,  Int. J. Mod. Phys.
			{\bf A12} (1989) 2861-2896
\bibitem{zajc}           W. A. Zajc, in NATO ASI Series {\bf B303}, p. 435
			(Plenum Press, 1993, ed. by H. Gutbrod and J.
			Rafelski)
\bibitem{padu}          Y. Hama and S. S. Padula,
			Phys. Rev. {\bf D37} (1988) 3237;
			S. S. Padula and M. Gyulassy, Nucl. Phys.
			{\bf B339} (1990) 378
\bibitem{lutp}          T. Cs\"org\H o and S. Pratt,
			{\bf KFKI-1991-28/A}, p. 75
\bibitem{sinyukov}      A. Makhlin and Y. Sinyukov,
			Z. Phys. {\bf C39}, (1988) 69
\bibitem{1d}            T. Cs\"org\H o, Phys. Lett. {\bf B347} (1995) 354-360
\bibitem{sinyu93}       Yu. Sinyukov, Nucl. Phys. {\bf A566} (1994) 589c-592c
\bibitem{hhm:te}        Yu. Sinyukov, in Proc. HHM:TE (Divonne,
			June 1994, Plenum Press, J. Rafelski et al, eds.)
\bibitem{uli}           S. Chapman, P. Scotto and U. Heinz,
			hep-ph/9409349, Heavy Ion Physics {\bf  1} (1995) 1;
			S. Chapman, P. Scotto and U. Heinz, hep-ph/9408207,
			Phys. Rev. Lett. {\bf 74} (1995) 4400
\bibitem{nr}            T. Cs\"org\H o, B. L\"orstad and J. Zim\'anyi,
			 Phys. Lett. {\bf B338} (1994) 134-140
\bibitem{schneder}      E. Schnedermann, J. Sollfrank and U. Heinz,
			Phys. Rev. {\bf C48} (1994) 2462
\bibitem{pratt_csorgo}  S. Pratt, T. Cs\"org\H o, J. Zim\'anyi, Phys. Rev.
			{\bf C42} (1990) 2646
\bibitem{gyulassy}      M. Gyulassy, S. K. Kaufmann and L. W. Wilson,
			Phys. Rev. {\bf C20} (1979) 2267;
			S. Padula, M. Gyulassy and S. Gavin,
			Nucl. Phys. {\bf B329} (1990) 203
\bibitem{swig}          S. Chapman and U. Heinz,
			Phys. Lett. {\bf B340} (1994) 250
\bibitem{bertsch}         G. F. Bertsch, Nucl. Phys. {\bf A498} (1989) 173c
\bibitem{htau}           T. Cs\"org\H o and J. Zim\'anyi,
			Nucl. Phys. {\bf A517} (1990) 588-598
\bibitem{lukacs}        E. Lukacs, {\it Characteristic Functions},
			(Charles Griffin and Co. Ltd, London, 1964);
			E. Lukacs and R. G. Laha,
			{\it Applications of Characteristic Functions},
			(Charles Griffin and Co. Ltd, London, 1964)
\bibitem{chapman_95}    S. Chapman, P. Scotto and U. Heinz,
			Nucl. Phys. {\bf A590} (1995) 449c
\bibitem{LSPS}          S. Chapman, J. R. Nix and U. Heinz,
			nucl-th/9505032
\bibitem{zsenya}        L. P. Csernai, J. Kapusta, Gy. Kluge and E. E.
			Zabrodin, Z. Phys. {\bf C58} (1993) 453-460
\bibitem{akkelin}       S. V. Akkelin and Yu. M. Sinyukov,
			preprint ITP-63-94E
\bibitem{ave}           V. A. Averchenkov, A. N. Makhlin and Yu. M. Sinyukov,
			Yad. Phys. {\bf 46} (1987) 1525-1234;
			Sov. J. Nucl. Phys. {\bf 46} (1987) p. 905
\bibitem{qm95}          T. Cs\"org\H o and B. L\"orstad,
			hep-ph/9503494,
			Nucl. Phys. {\bf A590} (1995) 465c
\bibitem{TD}            T. Cs\"org\H o and L. P. Csernai,
			Phys. Lett. {\bf B333} (1994) 494-499
\bibitem{halo}          T. Cs\"org\H o, B. L\"orstad and J. Zim\'anyi,
			LUNFD6-NFFL-7088, hep-ph/9411307
\bibitem{lpte}          T. Cs\"org\H o, LUNFD6-NFFL-7092-1994,
			 hep-ph/9412323
\bibitem{wiedemann}     U. A. Wiedemann, P. Scotto and U. Heinz,
			TRP-95-12,  nucl-th/9508040
\bibitem{na49}          T. Alber et al, NA49 collaboration,
			Nucl. Phys. {\bf A590} (1995) 453c
\bibitem{glasgow}       B. L\"orstad, NA44 Collaboration,
			Proc. 27-th Int. Conf. on High Energy
			Physics, Glasgow, July 20-27 1994 (IOP, 1994) p. 513
\end{thebibliography}
\end{document}